\documentclass[aps,prd,superscriptaddress,eqsecnum,amsfonts,showpacs,epsf]{revtex4}
\usepackage{epsfig}

\newcommand{\be}{\begin{equation}}
\newcommand{\ee}{\end{equation}}
\newcommand{\bea}{\begin{eqnarray}}
\newcommand{\eea}{\end{eqnarray}}

\newcommand{\nn}{\nonumber}
\newcommand{\ep}{i\epsilon}

\begin{document}


\title{Intriguing solutions of the Bethe-Salpeter equation for  radially excited pseudoscalar charmonia}

\author{V. \v{S}auli}

\email{sauli@ujf.cas.cz}
\affiliation{Department of Theoretical Physics, Institute of Nuclear Physics Rez near Prague, CAS, Czech Republic  }

\begin{abstract}

The purpose of this paper is twofold. The first purpose is to find a fully Poincare invariant solution  of the Bethe-Salpeter equation for 
excited quarkonia, however, the second,  in fact, major focus is on the relevance of the space-time metric choice and its imapact on the 
correct description of the ground and  all excited states. For the first time, we compare  BSE solutions defined independently with  Euclidean and Minkowski metric. For this purpose, the BSE is conventionally defined and solved in Euclidean space with two versions of the propagator
: the bare propagator  and the confined form of the quark propagator with complex conjugated poles. In both considered cases, there is  unexpected doubling of the spectrum, when comparing to the experiments  as well as to  the solutions  of  the Schrodinger equation.
The quark propagator with complex conjugated singularities allows us to find the BSE solution directly in Minkowski momentum space as well. We find the Minkowski space solution for confining theories is not only numerically accessible, but provides a reliable, albeit not yet completely satisfactory, description of the ground and  excited meson states.

\end{abstract}

\maketitle


%
\section{Introduction}

Excited meson spectroscopy is a keystone experimental output, which
is essential for understanding  quark-antiquark interaction. 
The dynamics of hadrons  is dominantly  driven by the solely known strongly interacting quantum field theory-
quantum chromodynamics.  The  confining forces are responsible for the absence of colored hadron constituents and simultaneously they  lead  to the spectra of ground state mesons and excited resonances. Because of the strong interaction at large distances $r\simeq 1/{\Lambda_{QCD}}$, the use of Schrodinger equation  becomes meaningless for  light hadrons. Unlike to the potential quantum-mechanical description, the excited states are not described by an orthogonal wave function in quantum field theory.
   
More precisely, within the model of logarithmic confinement \cite{QUIROS1979,EIMARO2007}, one has $<v^2>=0.24$ for quark of mass $m_c=1.5$GeV bounded inside of $1S$ charmonia. Thus,  charmonia lie on the borderline 
of a nonrelativistic description applicability (while this is less urgent for bottomonia, where one has $<v^2>=0.08$ for the ground state). 
In fact, one  requires the use of quantum field description in order to maintain Lorentz invariance, however most of the results (e.g. spectroscopy)  still should be consistent with usual quantum-mechanical machinery. In this respect, the physics of charmonia represent exciting theoretical laboratory where various methods can coexist.

 Recently in the CLEO \cite{CLEO} and Belle experiments progress in the experimental determination of electromagnetic transitions has provided new data, which  should be confronted with fully Poincare invariant description. Recall that  in naive quantum mechanical quarkonium picture $\psi(nS)\rightarrow \eta(n^{'}S) \gamma$ transitions should vanishes in zero recoil $\gamma$ for different $n,n^{'}$ due to the orthogonality of radial wave functions (see for instance \cite{Eichten1975, Eichten1980}). This is not the case of relativistic treatment in Bethe-Salpeter equation (BSE) framework, where this so called "hindered" transitions are not expected to be vanishing even for small photon energy. Also, a lot is known from  heavy quarkonia production in $e^{+}e^{-}$ anihilations, where vector quarkonia are straightforwardly produced. The BABAR \cite{BABAR}, Belle and BESS experiments continue collection of various meson experimental data.

In  quantum field theory the two body bound state is described by 
the Bethe-Salpeter (BS) vertex function or, equivalently, by 
the BS amplitudes. Both of  them are solutions of the 
corresponding  covariant four-dimensional BSE, which has definite structure dictated by the total spin of the meson. 
When generalizing recent various  quantum-mechanical models to quantum field theoretical approach,  one is faced with the solutions that do not exist in the nonrelativistic limit.  
Some of these states, the so called ghosts, are believed to be an artefact of inconsistent approximations or even illness of the theory at all \cite{ALKOFER} and we will not  discuss these states in this paper. However some of these additional states can not be apriory excluded as they have the same symmetry as expected -physical- ones. For  partial simplicity of related BSE structure we have started with pseudoscalar charmonia in this paper, recalling that the heaviest candidate for excited pseudoscalar $\eta_c$ meson was $X(3940)$ observed by Belle \cite{Belle} in double charmonium production is very likely $\eta(3S)$ state.

The  experimental data accompanied by quantum mechanical hints of nonrelativistic QCD limit can 
guide the systematics of hadronic excited states obtained in various different frameworks. Actually we suppose that the charmonium is ideal system, which can be used when concerning and discriminating the properties of strong coupling models defined in different metrics.  
 While it is commonly assumed that QCD formulated within the Minkowski metric and in  Euclidean space are equivalent,
we will regard  Euclidean QCD and Minkowski QCD as  independent theories. 
Let us stress, that having no comparisons in the literature,  the equivalence of the Minkowski and Euclidean formulation is very 
formal. It is believed or assumed that  the solution in the timelike Minkowski space can be uniquely obtained from the Euclidean  by analytical continuation.  While this  procedure is well defined in  perturbation theory calculus, it is not straightforward in confining theory like QCD. Actually, it  has been known  for almost a quarter century, that the (light) quark propagator exhibits complex conjugated branch points \cite{STACAH1990} and,  very similarly, it is assumed for gluons as well \cite{STINGL1,STINGL2}.  Such presence of complex conjugated singularities implies inequivalence of the original Minkowski theory and the theory defined in the Euclidean space. Recall that the usual Wick rotation does not provide correct Euclidean-Minkowski continuation in this case. In order to provide first insights, we start with the historically more usual  Euclidean formulation of BSE for charmonia,and we perform a realistic non-perturbative calculation using the Minkowski metric as the defining one. Assuming  the Euclidean theory offers  good approximation, we use the quark propagator with complex conjugated singularities  for the Minkowski space calculation as well. 
As we will see, contrary to popular belief,  the obtained numerical spectrum  prefers the Minkowski metric.

In the recent paper \cite{JA},  vector charmonia have been considered in the approximation with one dominant component. In the present paper, we continue  to study for pseudoscalar $c\bar{c}$ mesons by using the BSE model. 
In the sections II and III, the Euclidean BSE for all components completely included  is discussed and the results are presented for an arbitrary excited state for the first time. The states with masses above $3.9$GeV are  predictions  of our presented  BSE model.
Section IV  is devoted to the novel Minkowski space-time treatment of the BSE for mesons.

\section{BSE for Pseudoscalar Charmonium in Euclidean space}

The BSE for the pseudoscalar meson in Minkowski space reads

\bea \label{BSEcelkove}
\Gamma(q,P)=-i\int \frac{d^4k}{(2\pi)^4}
\left[S(q-P/2)\Gamma(p,q)S(q+P/2)\right]^{i,j} V(k,q,P)_{i,k,l,j} \, \, , 
\eea
where latin letters $i,j...$ represent Dirac indices. Explicitly for the pseudoscalar, we have
\be \label{general}
\Gamma_P(q,P)=\gamma_5\left(A(q,P)+ \not P C(q,P)+ \not q B(q,P)+[\not q,\not P] D(q,P)\right) \, \, ,
\ee
   
Alternatively, the BSE vertex function can be replaced by a (more singular) BS wave function $\chi$
through its definition
\be  
\Gamma(q,P)=S^{-1}(q-P/2)\chi(q,P)S^{-1}(q+P/2) \, \, , 
\ee
which is usually considered when performing the nonrelativistic limit.  The interaction kernel is given by the infinite sum of the two-particle irreducible quark-antiquark scattering graphs in a color-anticolor channel.

According to the original idea of Ref. \cite{Wilson:1974}, confinement in the heavy flavor hadron sector is typically  associated with a linearly   rising potential between constituents \cite{QUIROS1979,Eichten1978}. The spin degeneracy observed in meson spectrum  tells us that the  confining part of the interaction should be largely spin independent- it should be a  Lorentz scalar.  Euclidean approximation ofthe lattice calculations leads to various predictions for the potential between heavy quark-antiquark states.  For instance, Coulomb gauge potential rises linearly with the slope larger than required by an approximate Regge trajectory of light mesons, while all interpolating gauges between Landau and Coulomb gauge produce potential which is bounded from above \cite{SUGANUMA}. Also in Refs. \cite{DICHAQI,Licha2009,Zhang,Chao1,Chao2,Vento2011}, it has been found that the meson spectroscopy is better described  by a "confining"  potential which is bounded from above.  Irrespective of the gauge, the flatness of linearly rising potential  arises from the string breaking scenario: including light quarks into the game then the creation of the quark-antiquark pairs, i.e. pions and other light mesons are energetically favorable. Such screening is in agreement with observed high radially excited heavy mesons- without any doubt, the spectrum deviates from the linear Regge trajectory.  According to nonrelativistic quantum mechanic predictions, the relatively small hyperfine and fine-structure splitting in quarkonium levels is due to the Lorentz vector part of the interquark interaction \cite{Appelquist1978,Novikov1978,Kwong1987,Brambilla2004}, which should be added for a more accurate description. Following the arguments stated here  the dominant part of $V$ should be a Lorentz scalar, and further vectorial interaction is naturally assumed in QCD (for a quarkonia mainstream, see Ref. \cite{QUARKONIA}).  The interaction  phenomenologically chosen here, thus, reads
\bea 
V(k,q,P)_{iklj}&=&1_{ik}1_{lj}V_s+\gamma_{\mu,ik}\gamma^{\mu}_{lj}V_v \, \, ,
\nn \\
V_s&=&\frac{\kappa_s}{(q^2-\mu_s^2)^2} \, \,
\nn \\
V_v&=&\frac{g^2}{q^2-\mu_v^2} \, \, .
\eea
Note the scalar part  corresponds with the  negative  exponential potential in the nonrelativistic limit, $V_s({\bf r}) \simeq -e^{-r\mu_s}$.

The color  Coulomb potential is usually taken  due to  the exchange of the  gluon. It should reflect  the freezing of the effective coupling in the infrared related with  the gluon propagator suppression due to the soft gluon mass generation \cite{Mendes,PBC1,PBC2} through the Yang-Mills Schwinger mechanism. Here, we simplify and approximate $V_v$  by a constant coupling  and constant gluon mass of expected  size $\mu_v \simeq \Lambda_{{\mbox QCD}}$.  Further, from the string  models, it is well known that the string tension changes with the scale as well. At least, it is considerably larger for the charmonium than for the bottomonium. Here we found advantageous to consider such running  already for different charmonia, and we simply separate the $1S$ state from the others in the following way:
\be
\kappa_s=\kappa\left[\theta(P^2-M^2(1s))+\delta\theta(-P^2+M^2(1s))\right]  \, 
\ee  
where numerically, $\delta=0.9$.

We formally start at Minkowski space and  transform 16 component of BSE into the coupled set of four integral equations for 
four components $A,B,C,D$. The Wick rotation and further details are discussed in the appendices.

heeeeeeeeeerrrrrrrrrrrrrre is the end

\section{Numerical method and results}

Independent of the details of the model, when solving BSEs in the Euclidean space, we  face  a rather intriguing problem,
which we call the doubling of  excited states. 
 With the exception of the ground state (and possibly, the 2s state) there are 2 times more excitations than in the nonrelativistic limit  (for the nonrelativistic limit, see, for instance \cite{Licha2009,Chao1,Chao2,Vento2011}). We have found that the doubling is numerically  stable, and we argue that this is a rigid property of the BSE numerical solution for heavy quarkonia in the Euclidean space.

Let us discuss the doubling phenomena in more detail. It is a matter of the fact that instead of one excited state, we observe an emergence of  two 
with very similar BSE wave functions.
The appropriate numerical procedure is described in  Appendix B with  a typical result  shown in Figs. 6 and 7. 
As it is explained in the Appendix B, we introduce the auxiliary eigenvalue function $\lambda$ in a way that whenever
  $\lambda $ crosses the unit, we get a bound state.  As mentioned, the  doublets are characterized by an almost identical vertex function, hence, they are  physically  distinguishable only by a slightly different $M$ . Numerically, they have been found as a neighborhood energy levels with the opposite derivative $d\lambda/dM$. 
As we will discuss below, the appearance of doubling  does not depend on whether the free quark propagators approximation or some confined form for the quark propagator is used. Actually, for a reasonable large value of the imaginary part of the complex conjugated mass term, the spectrum of the model with confine quarks
is quite similar in both Euclidean models under consideration. Varying the imaginary part of the pole position can shrink the energy gap between the doublet pair; however, it leads to an unacceptable changes of the charmonium spectrum before the doubling is eliminated.  
We infer that  the phenomena of energy doubling is not related  solely  with analytical properties of quark propagators, but it is a  consequence of the use of Euclidean metric as a definite one.

To get rid of this problem we solve the BSE with two approximations of the quark propagator.  The free quark propagator
in the first:
\be \label{free}
S(k)=\frac{\not k+ m_c}{k^2-m_c^2+\ep} \, ,
\ee  
where $m_c\simeq 1.5 $GeV is the real "constituent" charm quark mass.

A confined type of quark propagator was used in the second approximation of the BSE in the Euclidean space.     
The charmonium quark propagator has been approximated by the following formula
\be \label{minkprop0}
S(k)=\left[\not k+ m\right]\frac{k^2-m_c^2}{(k^2-m_c^2)^2+\delta^4} \, ,
\ee
which is written in Minkowski space and where $m_c, \delta $ are real parameters. 
Such form of the propagator is in accordance with confinement \cite{maris1995,STINGL1,STINGL2,DUDAL1,DUDAL2,lat1,lat2,CLOROB2013},
and for a quark with momentum ${\bf k}\simeq \delta$, the parameter $\delta$ can be interpreted as minimal wavelength \cite{SHROCK}. It does not have a real pole corresponding to the free particle solution, but, instead,  there are complex conjugated singularities, as wassuggested almost 25 years ago in \cite{STACAH1990}.

The change of the free quark propagator into the form (\ref{minkprop0}) is equivalent to replace  the double Euclidean propagator $G_2$ in (\ref{eub})
in a way  such that $G_2\rightarrow G_{new}$, where
\be \label{technote}
G_{new}=\frac{I_{free}}{I_{free}^2+\delta^8+\delta^4\left[(k^2_E-\frac{P^2}{4}+m_c^2)^2-k_4^2 P^2+\lambda_{inf}\right]} \, 
\ee 
and where $I_{free}$ corresponds with the inverse of the original $G_2$, i.e.,
\be
I_{free}=(k^2_E-\frac{P^2}{4}+m_c^2)^2+k_4^2 P^2 \, \, .
\ee

Nowadays, up to the first two, the excited pseudoscalars have not been accessible in the  experiments. We remind here, that the  highest  mass candidate suggested  is $\eta(3s)$ observed in the Belle experiment in 2007 and hopefully will be confirmed in the future experiments in the recently built FAIR facilities. 
It is well known that the masses of $0^{-+}$  excited states  should be approximately degenerated with $S-states$ vectors charmonia. The five or six s-wave dominated  $\psi$ charmonium  vectors  are recently known with masses  spread in the energy range $3-4.5(5)$GeV \cite{RUPP}. No doubling is observed there and thus we do not expect doubling in the other channels as well. 
Remind here, the BSE vertex function  depends on two variables :$P.q$ and $q^2$. Alternatively and conventionally, vertex function can be visualize as a two dimensional manifold above $q_4$ and ${\bf q}$ plane. 
The BSE solution with free quark propagators for $\eta(1S)$ is shown at Fig. 1 for a fixed $p_4=0.25$ GeV. The doublets have been actually searched by  by comparison of the  vertex functions.  Examples of a few neighbor excitations  are shown  in Figs. 2-4. The order of the energy levels are used to  label the solutions (from low to high), and  the letter represents the components of the BSE vertex function [according to Eq. (\ref{general}].

\begin{figure}[t]
\begin{center}
\centerline{  \mbox{\psfig{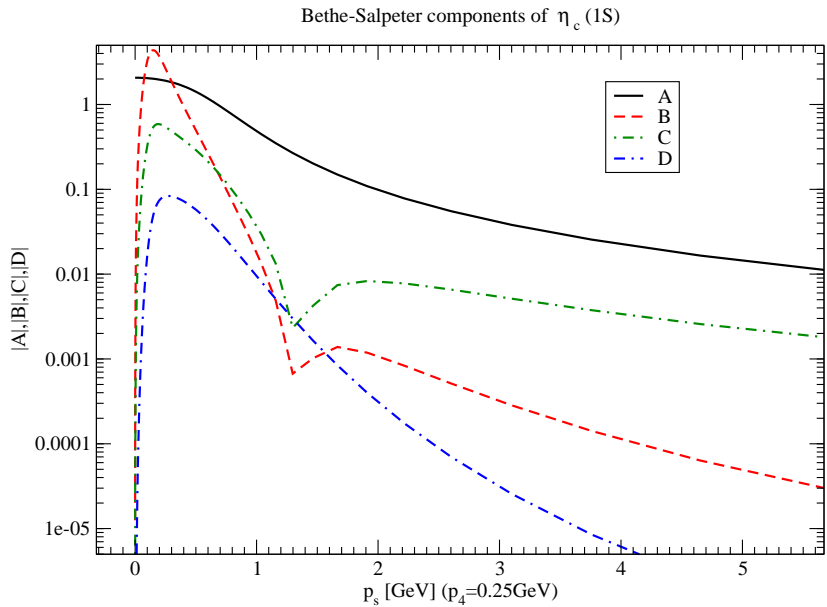}} }
\label{regge1}
\caption{Absolute value of the $A,\beta,C,D$ components of $\eta(1s)$ state. Note the large value of the function $ \beta$.   }
\end{center}%
\end{figure}
\begin{figure}[t]
\begin{center}
\centerline{  \mbox{\psfig{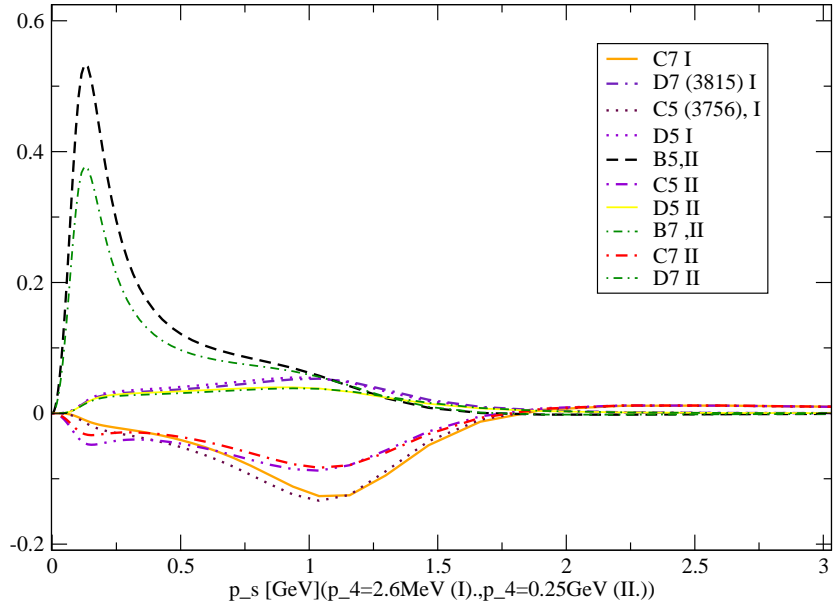}} }
\label{regge2}
\caption{Dependence of components $B,C,D$ on ${\bf k}$ of two  excited states shown for two  slices with given  $k_4=2.6MeV $ (I) and $k_4=0.25 GeV$ (II). }
\end{center}%
\end{figure}
\begin{figure}[t]
\begin{center}
\centerline{  \mbox{\psfig{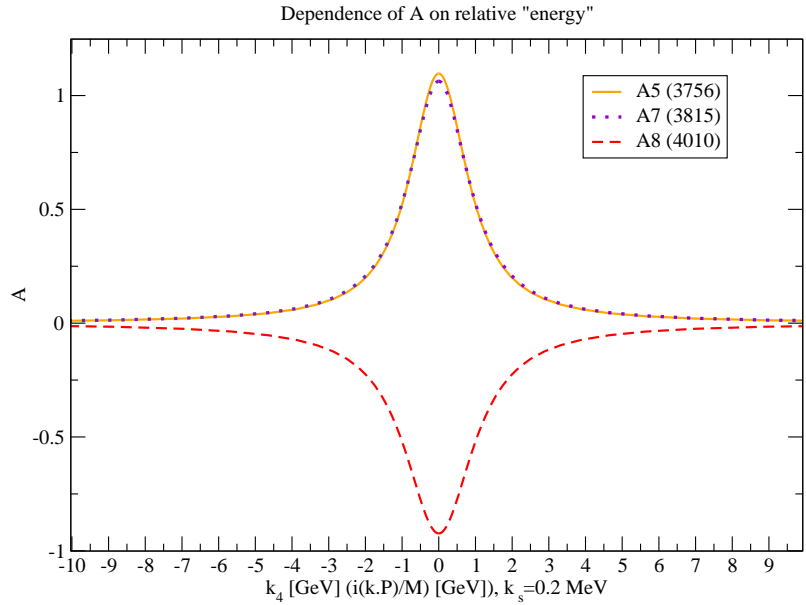}} }
\label{regge3}
\caption{Dependence of dominant component $A$ of three neighbor excited states shown for fixed threemomentum ${\bf{k}}=0.2MeV $. Vertex functions become wider for higher mass of $\eta_c$ meson. Quarks "are more off-shell" in higher excited mesons for the model with a free propagators.}
\end{center}%
\end{figure}
\begin{figure}[t]
\begin{center}
\centerline{  \mbox{\psfig{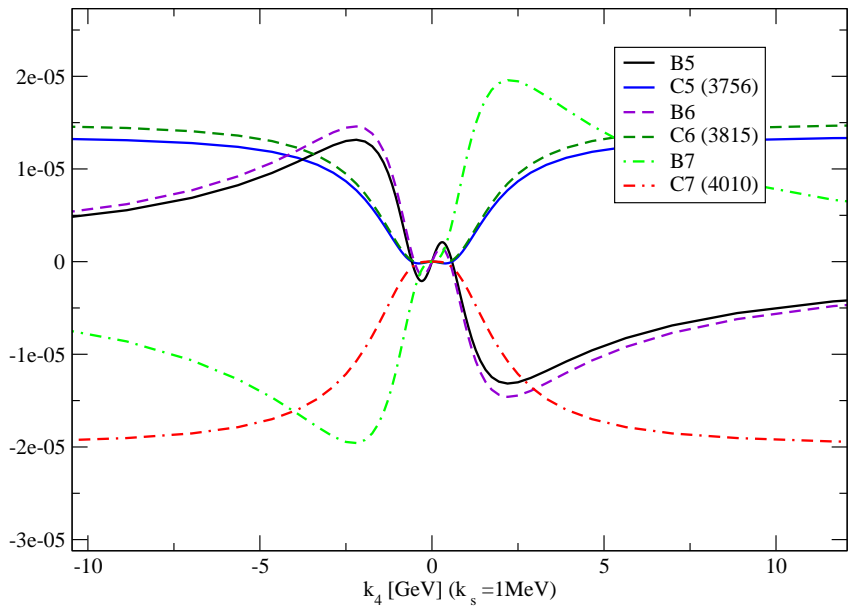}} }
\label{regge4}
\caption{Dependence of B and C components of three excited states shown for fixed threemomentum ${\bf{k}}=0.2$MeV . Levels labeled by B(C)5 (fifth energy excited state) and B(C)6 (sixth excited state) belong to the neighboorhoodhred levels, they have  approximately identical vertex functions.
 They have same nodes (odd function B has three nodes, while C functions have two  nodes and zero minimum at beginning, next two higher levels have only one zero, and only the seventh is shown for better visuality). }
\end{center}%
\end{figure}
\begin{figure}[t]
\begin{center}
\centerline{  \mbox{\psfig{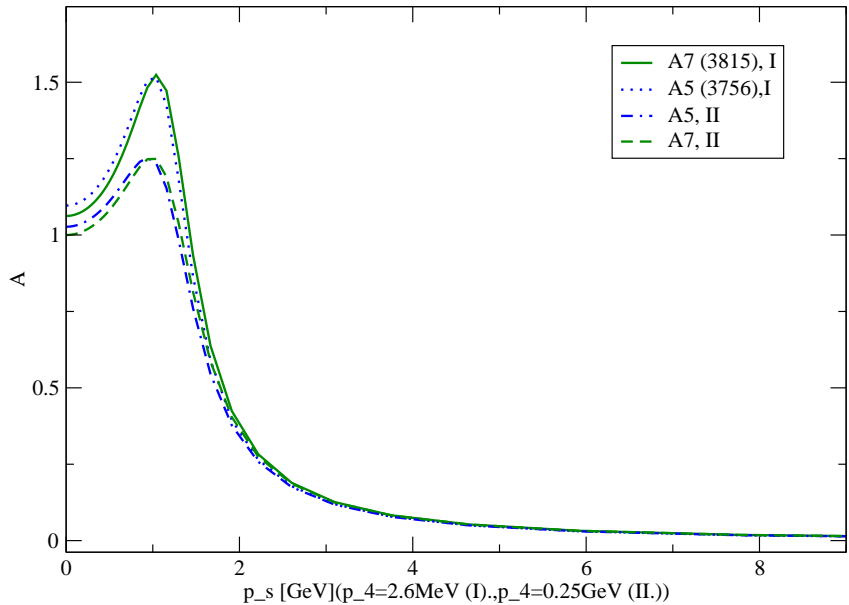}} }
\label{reggex}
\caption{Dependence of dominant component $A$ of two  excited states shown for two  fixed values  $k_4=2.6$MeV  and $k_4=0.25$ GeV. The function $A$  has no nodes for Euclidean relative momentum $k$.}
\end{center}%
\end{figure}

For a rough estimate of the spectra in the case of BSE with free propagators,  we  adjusted charm quark pole mass to be
 $m_c=1.5$ GeV and we were searching for the charmonium spectra for different parameters
$\kappa(P),\mu$. The strength of the effective vectorial interaction $\alpha_s=g^2/(4\pi)$  has been adjusted in order to adjust the intercept to the experimentally known value $\eta(2s)-\eta_c(1s)=660$GeV. According to the nonrelativistic calculations we  were varying the parameters $\kappa$ between $2-4 GeV^2$ and $\mu=0.1-0.5$GeV. At the end the the numerical data have been rescaled in order to get correct value of the ground state exactly. The resulting search is presented  in the Table 1. , where the all pairs belonging to the single nonrelativistic counterpartner are matched. 
In the case of free quark propagator approximation  all states are realized above naive quark threshold.

\begin{table}[b]
\begin{center}
\small{
\begin{tabular}{|c|c|c|c|} \hline \hline $m_c=1500$ &
$m_c=1442$  & $< >$ & Exp.  \\
\hline \hline
3100 & 2980 & 2980 & 2980 (1s)\\
\hline
3785   & 3638 & 3638 & 3638 (2s)\\
\hline
3940  & 3787 & &\\
3990 & 3835 &  3811 & 3940* (3s) \\
\hline
4160 & 4000 &  &\\
4235 & 4071 & 4036 & - (4s)\\
\hline 
4430 & 4259 & &\\
4535 & 4360 & 4310 & -(5s)\\
\hline
4790 & 4373 &  &\\
4925 & 4734 & 4554 & -(6s)\\
\hline
5270 & 5066 &  &\\
5435 & 5224 & 5145 &-(7s)\\
\hline \hline
\end{tabular}}
\caption[99]{Conventional BSE solutions for $\eta_c(nS)$ for the model with the free quark propagator. We use conventional quantum mechanical assignment $nS$ in order to label states that we expect in nonrelativistic or "instantaneous" approximations. The first column  represent actual numerical solution in units where $m_c=1.5GeV, \kappa=2.849 GeV^2$, in the second column the experimental value of $\eta(1S)$ has been used to scale other levels. The doubling appears for the  states $n>2$, and the energy doublets are identified by comparison of vertex functions (e.g. by number of nodes in $B,C,D$ functions).  After rescaling we produce experimentally known $\eta(2S)$.  For other states -to make levels meaningfully comparable with quantum mechanical labeling- the masses of energy levels are averaged for given energy doublets.  $\alpha_s=0.07407$ . *Belle observed X(3940) in $e^+e^- \rightarrow J/\psi +X$, for the interpretation see \cite{EIMARO2007}.  
\label{tab_bse}}
\end{center}
\end{table}

For the purpose of numerical study of BSE with confinement incorporated through  Eq. (\ref{minkprop0}),  we have used the same parameters as in the case for BSE with free propagators, e.g. the real part of the
complex poles is  $m_c=1.5$GeV. The  parameter $\delta$ characterizes the splitting of complex conjugated poles
 and we present the results for two values $\delta=0.6$GeV and $\delta=1.0V$GeV graphically in Fig. 6.
The situation for highly excited states is quite similar to the case with unconfined quarks. However, there are new states arising  where just only  one -$\eta(2)$- should exists, to be more precise there are two new (4 because of doubling) states for $\delta=0.6 Gev$ and three (six) additional states for $\delta=1 GeV$. While the gaps between the partners of  energy doublet  slightly and tentatively shrink, the appearance of  new states makes any value of $\delta$ inappropriate for a realistic Euclidean model.     

Another numerical observation is that the vectorial interaction   must be largely suppressed  in our model, which gives typical value estimate $\alpha_s\simeq 0.07-0.1 $. This is several times smaller then the expected value of the strong coupling in nonrelativistic models, where, typically, one gets $\alpha_s(m_c)=0.5$. It is difficult to trace the source of such large suppression. We have explicitly checked that the introduction of the running strong coupling does not explain its suppression in presented model. A more complete understanding of this problem remains for a future  study. 
For a ground state mesons, it is well known that the function $A$ largely dominates. We have found that this is the case for excited mesons as well. This  is apparent from the Fig. \ref{lambda} , where the numerical results are graphically compared for BSE with free propagator.

\begin{figure}[t]
\begin{center}
\centerline{  \mbox{\psfig{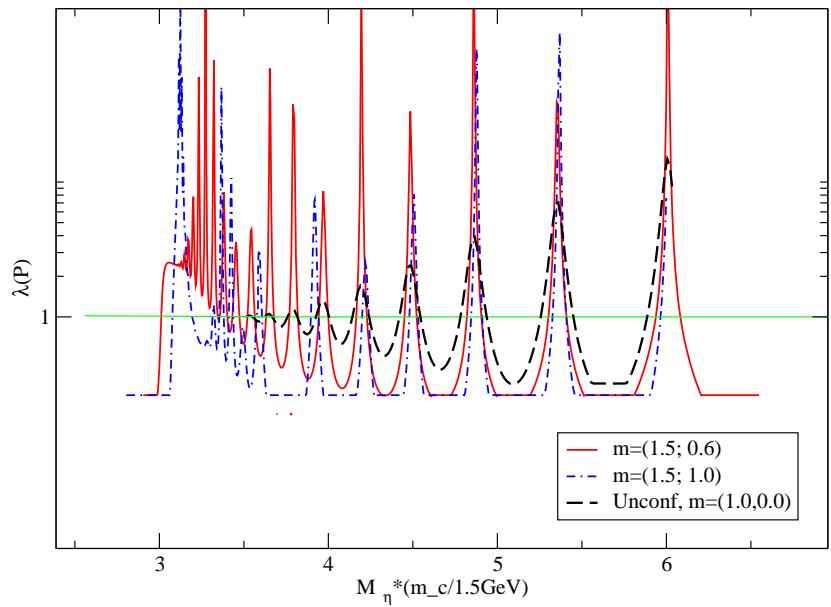}} }
\label{lama1}
\caption{Comparison of BSE solution within various approximation, the solid line represents free quark propagator approximation with $m_c=1.5GeV$, the dashed line and dot-dashed line correspond with confined quark propagator characterized by two complex conjugated poles with
located at $m_c=1.5 GeV \pm i 0.6 GeV $ and  $m_c=1.5 GeV \pm i 1.0 GeV $ respectively. The BSE has solution whenever a curve crosses the horizontal solid line. Energy doublets corresponds with crosses, which cuts the upper parts of each curve separately.}
\end{center}%
\end{figure}

\section{BSE Minkowski space solution with confining propagators} 

 It is obvious that the propagator  (\ref{minkprop0})
is a regular function for a real Minkowski variable $k^2$. It allows us to consider and explicitly numerically solve the analogue problem in the Minkowski space directly. We completely avoid the use of Euclidean metric by using the confined form of the propagator as in the previous case, 
however, for  now, we stay in the Minkowski space-time.
We  will use the following convention 
$k^2=k_{\mu}k_{\nu}g^{\mu\nu}=k_0^2-k_1^2-k_2^2-k_3^2$, and we take for the quark propagator,      
\be \label{minkprop}
S(k)=f\left[\not k+ m\right]\frac{k^2-m_c^2}{(k^2-m_c^2)^2+\delta^4} \, ,
\ee
where, in  general,  $f$ is complex  function of the momenta, which, in accordance with the perturbation theory, must become a real valued function for a large spacelike $k$ (recall here that the Dyson-Schwinger equations for propagators are not directly numerically  soluble in momentum Minkowski space \cite{SAULISDE}  ). Here, in order to approximate its infrared behavior only, we simplify and take the following  constant phase approximation:
\be
f={\mbox exp}^{i\frac{\pi}{2}} \, .
\label{choice}
\ee

To make a comparison meaningful we keep the Lorentz structure  of the kernel as in  the previous Euclidean study, however in order to make the model   tractable, we change the double pole of the scalar kernel  to the double pair of complex conjugated poles while leaving the vectorial interaction in its original  form (without a Feynman epsilon), explicitly written as
\bea
V_s&=&\frac{\kappa}{(q^2-\mu_s^2)^2+\lambda_s^4} \, \, ,
\nn \\
V_v&=&\frac{g^2}{q^2-\mu_v^2} \, \, .
\eea

It is worth mentioning that the Minkowski space-time BSE represents well-defined numerical problem, and the convergence is further enforced  due to the asymptotic of the vertex function $\Gamma \simeq 1/q^2$ at large $q^2$ .
 In this way, the BSE turns out to be two-dimensional integral equation for two  real variables: two  scalars $k^2,k.P$ for given discrete $P^2=M^2$, or alternatively, for the off-shell relative energy $k_0$ and spacelike relative momentum ${\bf k}$.  According to our Euclidean space founding we keep only the component  $A$ as a reasonable approximation of the full BSE amplitude.
The BSE in Minkowski space  has been solved numerically in a very similar fashion as in the previous Euclidean case. Note also, that due to our simplifications,  the Minkowski  vertex function  remains real valued in the entire Minkowski space.

To specify the rest, we put  $\mu_s=\mu_v=\mu=0.25$GeV and take  $\lambda_s=2\mu$ for the purpose of our  numerical search. Having the double pole of $V_s$ changed to  the complex conjugated poles, the interaction strength is effectively suppressed. Within a given value of $\lambda_s$, one needs
to increase the scalar coupling strength. Numerically,  now we take $\kappa=5GeV^2$ and for the vector, we take $g^2=1.8849 $. Within $m_c=1.5GeV$ the spectrum we have found reads (without any further tuning):
$M(1s)=1.97 GeV; M(2s)=3.563GeV; M(3s)=4.163GeV; M(4s)=5.9 GeV; M(5s)=6.68GeV  $.   The doubling of energy levels in the numerical solution of the BSE has disappeared. The intercept is presented, although it is overestimated. We believe that  quantitative disagreement with the experiment will be further minimized after a further refinement of the parameters used in the presented model and/or after a further improvement of the approximation made.


\section{Conclusion}

The homogeneous BSE for excited states of pseudoscalar charmonia have been solved in the framework of BSE defined in Euclidean and Minkowski space-time
independently. A high number of excited states has been found  by the numerical solution of the BSE in  both spaces,  and  the numerics has been  described  in detail here. For those who are interested, the codes are available by an email request.  We have found that the excited states are overpopulated when calculated in the Euclidean space.  The excited states come in pairs with close masses and almost identical wave functions. This doubling phenomena is very likely an artifact of the Euclidean space approximation. In the Minkowski space-time  model with confining-type  of the quark propagators, the   doubling phenomena of the energy levels disappears. At the recent stage the Minkowski BSE  model  suffers by the  simplifications and approximations made, and it does not reproduce the experimental data precisely.   Nevertheless, it is reliable  enough to exhibit the Minkowski space metric can be used as a definite metric for nonperturbative QCD calculations as a phenomenological, if not a preferable then certainly an acceptable, choice.

\section{Appendix A: Projecting Euclidean BSE with the free fermion propagator}

After the projection by $\frac{1}{4}{\mbox Tr} \gamma_5$ we get in Minkowski space :

\bea  \label{acko}\label{ivanka}
A(q,P)&=&i\int\frac{d^4k}{(2\pi)^4} K_A  (V_S-4V_V)
\nn \\
K_A&=&S_V(-) S_V(+) \left[-A(k,P) \left(k^2-\frac{P^2}{4}\right)+2D(k,P) \left(k^2 P^2+(k.P)^2\right)\right]
\nn \\
&+& S_S(-) S_S(+) A(k,P)  +S_V(-) S_S(+) \left(k.P B(k,P)- \frac{P^2}{2} C(k,P)\right) \, \, ,
\eea
where we have introduced the shorthand notation $S(\pm)=S(k\pm P/2)$. 
Projecting the BSE by ${\mbox Tr} \frac{\not q \gamma_5}{4}$ one gets 
\bea \label{becko}
&&q^2 B(q,P)+q.P C(q,P)=i\int\frac{d^4 k}{(2\pi)^4} \left(q.P K_{B1}+q.k K_{B2}\right)  (V_S+2V_V)
\nn \\
K_{B1}&=& S_v(-) S_v(+) \left[\frac{k.P}{2} B(k,P)+
\left(k^2+\frac{P^2}{4}\right) C(k,P)\right]+  S_s(-) S_s(+) C(k,P)  - 4 k^2 S_v(-) S_s(+) D(k,P)
\nn \\
K_{B2}&=&-S_v(-) S_v(+)\left[2 k.P C(k,P)+\left(k^2+\frac{P^2}{4}\right) B(k,P)\right]+  S_s(-) S_s(+) B(k,P)
+ 4 k.P S_v(-) S_s(+) D(k,P)\, \, .
\eea
Projecting BSE by  $Tr \frac{\not P \gamma_5}{4}$ one gets in Minkowski space,
\bea \label{cecko}
&&q.P B(q,P)+C(q,P) P^2=i\int\frac{d^4 k}{(2\pi)^4} K_C  (V_S+2V_V)
\nn \\
K_C&=&-S_v(-) S_v(+) B(k,P)k.P(k^2-P^2/4)+  S_s(-) S_s(+) B(k,P)k.P
\nn \\
&-&S_v(-) S_v(+) C(k,P)\left[2(k.P)^2-k^2P^2-\frac{P^2}{4}\right] + S_s(-) S_s(+) C(k,P)P^2
- S_v(-)S_s(+) A(k,P) P^2 \, \, ,
\eea
 where we have dropped out the trivial term proportional to $D$ due to the identity 
$ S_V(-)S_V(+)-S_V(+)S_V(-)=0$. Finally the "equation for
$D$" reads 
\bea \label{decko}
&&4\left[(q.P)^2-q^2P^2\right]D(q,P)=i4\int\frac{d^4 k}{(2\pi)^4} K_D  (P.qP.k V_S-P^2 (k.q)V_S)
\nn \\
K_D&=&  S_V(-)S_V(+)\left[\frac{A(k,P)}{2}+(k^2 -P^2/4)D(k,P)\right]-S_V(-) S_S(+) C(k,P)+S_S(-) S_S(+)D(k,P) \, \, . 
\eea

 To find the solution numerically, it is convenient to  perform Wick rotation in the relative momenta, while keeping  $P^2$ real and positive. In the rest frame of quarkonium $P=(M,0,0,0)$ and after the continuation $q_o\rightarrow iq_4$ we get for the "Euclidean" Eq. (\ref{ivanka}):
\bea \label{eua}
A(q,P)&=&\int\frac{d^4 k_E}{(2\pi)^4} U_A G_2 (V_S-4V_V)
\nn \\
U_A&=&A(k,P)\left(k^2_E+\frac{P^2}{4}+m_c^2\right)+2D(k,P)\left(-k^2_E P^2+{(k_4 M)}^2\right)
+m_c k_4 M \beta(k,P) -m_c \frac{P^2}{2} C(k,P) \, ,
\eea
where  $A(q,P),U_A,...$ are the functions
of Euclidean  momentum $q_E=(iq_0,{\bf q})=(q_4,{\bf q})$ and Minkowski momentum $P$.  
From now on we restrict to the use of the free quark propagator
\be
S^{-1}(l)=\not l-m_c\, , 
\ee  
and introduce shorthand notation for the product of the propagators
\be
G_2=S_v(-)S_v(+) \, .
\ee
Thus,  we can simply write 
\be
S_s(-)S_s(+)=m_c^2 G_2\ \ \, \ \ S_v(-)S_s(+)=m_c G_2 \, ,
\ee
whose identities we will use from the next.

It is easy to find that $G_2$  is the product of two complex scalar propagators
, which for equal mass case, becomes purely real,
\be
G_2=\left[(k^2_E-\frac{P^2}{4}+m_c^2)^2+k_4^2 P^2+\lambda_{inf}\right]^{-1} \, \, ,
\ee
where we have introduced infrared  regulator $\lambda_{inf}<<m_c,M$ for a later numerical  purpose.
Furthermore, we have introduced new real function $\beta$ as
\be
\beta=iB \, ,
\ee
which reflects the fact that the function $B$ is odd in the variable $k.P=ik_4P$ (valid for a positive $C$ parity eigenstate).The functions $A,\beta,C$ and $D$ are then  manifestly real.  For the purpose of completeness, we write  the Euclidean  Eqs. (\ref{acko},\ref{becko},\ref{cecko},\ref{decko})  here
\bea  \label{eub}
q^2_E \beta(q,P)+q_4M C(q,P) &=&\int\frac{d^4 k_E}{(2\pi)^4} 
\left(q_4 M U_{B1}+q_E.k_E U_{B2}\right)  (V_S+2V_V) G_2
\nn \\
U_{B1}&=&\frac{k_4M}{2}\beta(k,P)+\left(- k_E^2+\frac{M^2}{4}+m_c^2\right) C(k,P)
+4m_c k^2_E D(k,P)
\nn \\
U_{B2}&=&\left(k_E^2-P^2/4+m_c^2\right) \beta(k,P) + 2k_4 M C(k,P)
- 4 k_4 M m_c D(k,P)
\eea

\bea \label{euc}
\left[q_4 \beta(q,P)+C(q,P)M \right]&=& \int\frac{d^4 k_E}{(2\pi)^4} U_C  (V_S+2V_V) G_2
\nn \\
U_C &=&  k_4(k_E^2+\frac{P^2}{4}+m_c^2)\beta(k,P)+ M\left[2k_4^2-k_E^2 +\frac{P^2}{4}+  m_c^2\right] C(k,P)
-M m_c A(k,P)
\eea

\bea \label{eud}
\left[-(q_4M)^2+q_E^2M^2\right]D(q,P)&=&\int\frac{d^4 k_E}{(2\pi)^4} U_D  \left(-M^2q_4k_4 V_S+P^2 (k_E.q_E)V_S\right)G_2
\nn \\
U_D&=&\frac{1}{2}A(k,P)-m_c C(k,P)+(-k^2_E -P^2/4+m_c^2)D(k,P)
\eea

In order to study excited states we keep full $k.P$ dependence  into account and do not perform any 3dimensional reduction which can scrutinize  a correct off-shell  behavior of the vertex function. For given bound state, characterized by the mass $M$ the four functions $A..D$
 depend only on the scalars $q.P$ and $q^2$ respectively, the first product  mix Minkowski and Euclidean metrics and become complex ($q.P --> iq_4M$ in CMS). Due to this fact we we prefer to leave $q_4$ integral variables separated from the spacelike part of the integration momentum. When reducing 4-dim integral equation we
therefore use the following angle integrals:
\bea
I_V^{[N]}=\int_{-1}^{1} d z\frac{(k_E.q_E)^{N-1}}{(k_E-q_E)^2+\mu^2} \, \, ;
\nn \\
I_S^{[N]}=\int_{-1}^{1} d z\frac{(k_E.q_E)^{N-1}}{[(k_E-q_E)^2+\mu^2]^2}\,\, ,
\eea
where $z$ is the  cosine of the angle between the space product of momenta $z=\frac{{\mbox{\bf k}}.{\mbox{\bf q}}}{|{\mbox{\bf k}}||{\mbox{\bf q}}|}, k.q=k_4q_4+{\mbox{\bf k}}.{\mbox{\bf q}}$. We need $N=1,2$ for our purpose, for  which cases the integrals read 
\bea
I_V^{[1]}&=&\frac{1}{2|k_s||q_s|} ln{\xi_+ \over \xi_-} \, ;
\nn \\
I_V^{[2]}&=&-1+\frac{k^2+q^2+\mu^2}{2}I_V^{[1]} \, ;
\nn \\
I_S^{[1]}&=&\frac{2}{\xi_+\xi_-} \, ;
\nn \\
I_S^{[2]}&=&-\frac{1}{2}I_V^{[1]}+\frac{k^2+q^2+\mu^2}{2}I_S^{[1]} \, ;
\nn
\eea
where
\be
\xi_{\pm}=k^2+q^2+\mu^2-2k_4q_4\pm 2\left|{\mbox{\bf k}}\right|\left|{\mbox{\bf q}}\right|\, .
\ee
 

Substituting $V_{s,v}$ and integrating over the variable $z$ we get 

\be \label{rovnice}
A(q,P)=\int_{-\infty}^{\infty} dk_4\int_{0}^{\infty}
\frac{d{\mbox{\bf k}} {\mbox{\bf k}}^2}{(2\pi)^3} U_A 
\left(\kappa I_S^{[1]}-4g^2I_V^{[1]}\right)G_2 
\nn
\ee
\be
q^2_E \beta(q,P)+q_4M C(q,P) =\int_{-\infty}^{\infty} dk_4\int_{0}^{\infty}
\frac{d{\mbox{\bf k}} {\mbox{\bf k}}^2}{(2\pi)^3}
\left[q_4 M U_{B1}\left(\kappa I_S^{[1]}+2g^2I_V^{[1]}\right)+ U_{B2}\left(\kappa I_S^{[2]}+2g^2 I_V^{[2]}\right)\right]   G_2 
\nn
\ee
\be
q_4M \beta(q,P)+C(q,P) M^2=\int_{-\infty}^{\infty} dk_4\int_{0}^{\infty}
\frac{d{\mbox{\bf k}} {\mbox{\bf k}}^2}{(2\pi)^3}
M U_C  (\kappa I_S^{[1]}+2g^2I_V^{[1]}) G_2
\nn
\ee
\be  \label{krabice}
\left[-(q_4M)^2+q_E^2M^2\right]D(q,P)=\int_{-\infty}^{\infty} dk_4\int_{0}^{\infty}
\frac{d{\mbox{\bf k}}{\mbox{\bf k}}^2}{(2\pi)^3} U_D  (-M^2q_4k_4\kappa I_S^{[1]} +P^2 \kappa I_S^{[2]})G_2 \, .
\ee

\section{Appendix B- Numerics}

In this appendix we write down the BSE in the form which has been actually used in our numerical solution.
Contrary to quantum-mechanical approach, in quantum field theoretical framework  two  different  excitations are not described by an orthogonal BS functions.  Also the number of nodes in various component of BS wave function is not driven by any obvious rule. Thus, for instance the dominant component $A(p,P)$ remains nodeless for the all observed excited states.  As different components of the BS amplitude vary differently  on the $P.q$ variable, we do not explore the more or less conventional  expansion into the orthogonal polynomials, which loses its efficiency  when, as one expects, a relatively large number of polynomials could be necessary to distinguish correctly between two  excited states. Therefore instead of this, we  solve the full two dimensional integral equation by the method of simple iterations. 

For the purpose of numerical solution ,we discretize, and step by step we scan $P^2$ region of the total momenta.  Within the step of few $MeV$   we are looking for the solution of the BSE with given $P^2$. Performing several hundred iterations for each given value of $P^2$ we identify  the solutions as those for which difference between two consecutive iterations vanishes.

The BSE for bound states is a homogeneous integral equation and it satisfies canonical  normalization condition:

\be \label{norm}
\frac{1}{6}=\frac{d}{dP^2}{\mbox \bf Tr }\, i \int\frac{d^4k}{(2\pi)^4}
\bar{\Gamma_p}(k,Q)S(+)\Gamma_p(k,Q)S(-)+...
\ee
where $Q^2=M^2$ and the trace is taken over the Dirac matrices. The conjugated vertex
$\bar{\Gamma_p}(k,Q)=C\gamma^T\Gamma(-p,Q)C^{-1}$ with the charge conjugation operator $C$ and three dots represent terms with derivatives of the kernel with respect to $P$ . The normalization condition (\ref{norm}) must be  necessarily considered when bound state transitions are considered. On the other hand the efficiency of numerical procedure is enforced by  the use of the auxiliary normalization which is called at each step during the iteration process.  
 For this purpose we implement the  auxiliary function $\lambda(P)$ and solve numerically the equations (\ref{rovnice}) with $\lambda(P)$ implemented in.  The physical normalization condition can be applied afterwords.

We found it was suited to rewrite Eqs. (\ref{rovnice}-\ref{krabice}) in the following equivalent form:
\bea \label{BSEEnum}
A(p,P)&=&\lambda(P)\int_{-\infty}^{\infty} dk_4\int_{0}^{\infty}\frac{d{\bf k k^2}}{(2\pi)^3} U_A 
\left(\kappa I_S^{[1]}-4g^2I_V^{[1]}\right)G_2 
\nn \\      
D(q,P)&=&\lambda(P)\frac{-(q_4M)^2+q_E^2M^2}{\left[-(q_4M)^2+q_E^2M^2\right]^2+\epsilon}
\int_{-\infty}^{\infty} dk_4\int_{0}^{\infty}
\frac{d{\bf k k^2}}{(2\pi)^3} U_D G_2 (-M^2q_4k_4\kappa I_S^{[1]} +P^2 \kappa I_S^{[2]})G_2
\nn \\
\beta(q,P)&=&\lambda(P)\frac{-(q_4M)^2+q_E^2M^2}{\left[-(q_4M)^2+q_E^2M^2\right]^2+\epsilon}
{\mbox{(lhs. of Eq.)}}q^2 +{\mbox{(rhs. of Eq. (22))}}q_4M
\nn \\
C(q,P)&=&\lambda(P)\frac{-(q_4M)^2+q_E^2M^2}{\left[-(q_4M)^2+q_E^2M^2\right]^2+\epsilon}
{\mbox{(lhs. of Eq.)}}P^2 +{\mbox{(rhs. of Eq. (23))}}q_4M \, ,
\eea 
where $\epsilon $ is a small numerical regulator and where the  last two equations are  equivalent to the Eq. (22) and Eq. (23) in the limit $\epsilon\rightarrow 0$. 
The integrals have been discretized and the system of equations solved by iterations as usually: The left hand side of Eqs. represents the $i-th$ iteration step result when $i-1-th$ iteration has been used to evaluate lhs. of Eqs. system.  Starting with reasonable zeroth approximation then 300-400 iterations
have been used for each fixed value of $P^2$, which was enough to get obviously convergent solution.

Our main goal is the  implementation of  the function $\lambda(P)$  which when properly chosen can radically increase the efficiency of the numerical search.  
Among several  working possibilities that we have checked we mention two: firstly
the choice $\lambda^{-1}(P)=A(0,0)$ appears to work well in many cases.
Second,   we have taken 
\be
\lambda^{-1}(P)=\frac{1}{2}\int{d k_4} \int d {\bf k} \frac{\left[A_i(k,P)+A_{i+1}(k,P)\right]^2}{k_4^2+{\bf k}^2+m_c^2} \, ,
\ee
which has been actually used for the  evaluation of the  results in the  presented paper.  The solution of the BSE are identified whenever $\lambda=1$. Nontrivially, when $\lambda=1$, the difference between  consecutive iterations vanishes at the same time. 

\begin{figure}
\begin{center}
\centerline{  \mbox{\psfig{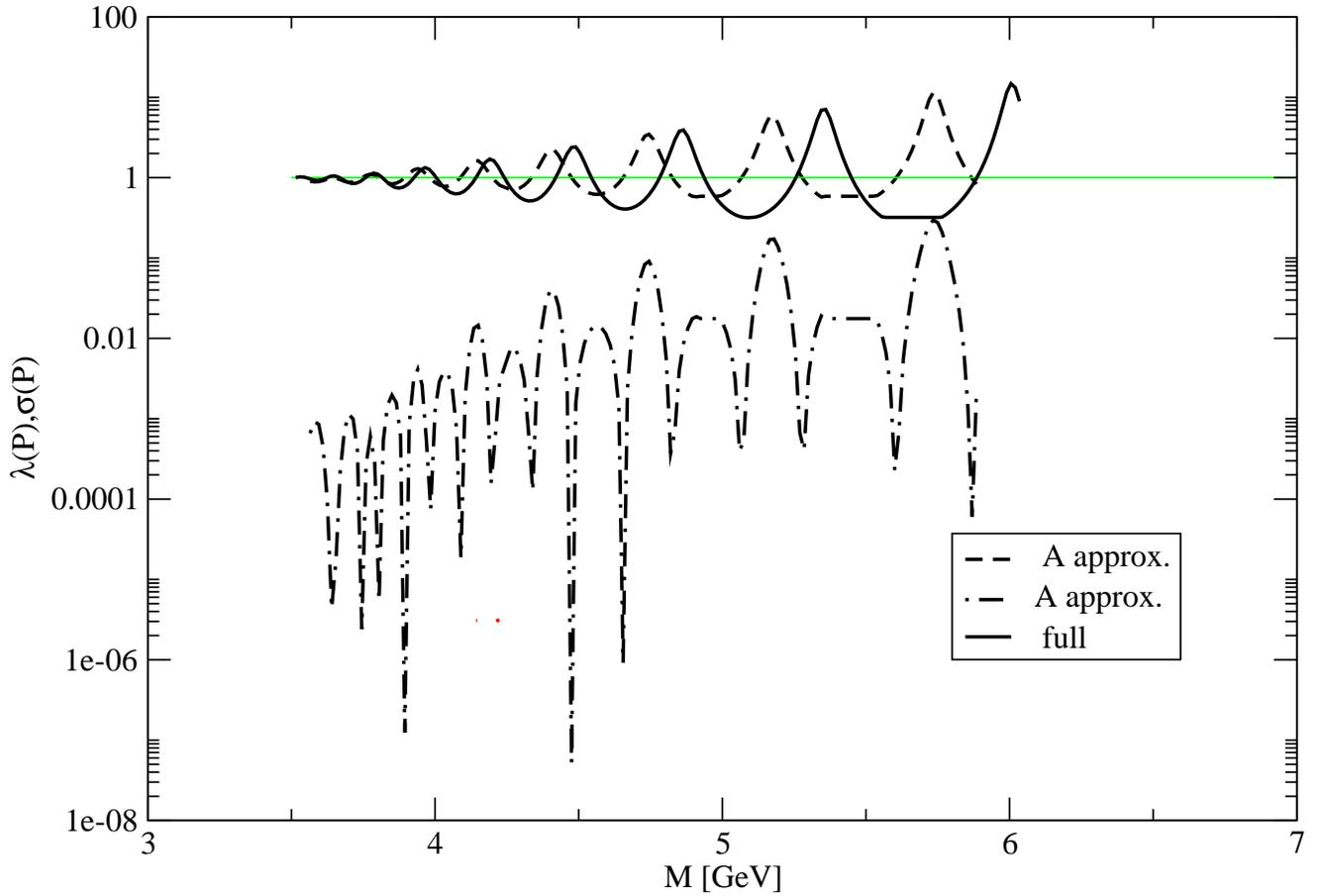}} }
\label{lambda}
\caption{Numerical search for $m_c=1.5 GeV, \kappa=2.5 GeV^2, \alpha_s=0.08, \mu_v=\mu_s=350MeV$ (solid line).
Dashed (dot-dashed) line represents $\lambda,(\sigma)$ in the approximation when only $A$ vertex function has been considered, in this case $m_c=1.5 GeV, \kappa=2.5 GeV^2, \alpha_s=0.0815,\mu_v=\mu_s=350MeV $.}
\end{center}
\end{figure}

The example of the solution of the BSE is shown in Fig.\ref{lambda} .  The function $\sigma$ is
the  weighted integral difference between two consecutive iterations evaluated through the following formula:
\be
\sigma(P)=\frac{1}{\lambda(P)}\int{d k_4} \int d{\mbox{\bf k}}
\frac{\left[A_i(k,P)-A_{i+1}(k,P)\right]^2}{k_4^2+{\mbox{\bf k}}^2+m_c^2} \, .
\ee

 To get a reasonable numerical error, we found that a relatively large number of integration points is required. Therefore, to speed up numerics we identify roughly the positions of the  bound states within small -$40*80$- number of mesh points. Consequently, the results has been improved within  -$88*176$-  points of discretized  integral variables ${\bf k},k_4$ .

\end{document}